\documentclass[twocolumn,showpacs,preprintnumbers,amsmath,amssymb,prl]{revtex4}

\pdfoutput=1

\usepackage{graphicx}
\usepackage{dcolumn}
\usepackage{bm}

\renewcommand{\vec}[1]{{\mathbf #1}}
\newcommand{\vecgreek}[1]{{\mbox{\boldmath$ #1$}}}

\begin{document}

\preprint{APS/123-QED}

\title{Photo-Emission of a Single-Electron Wave-Packet in a Strong Laser
Field}

\author{Justin Peatross$^{\dag}$}
\email{peat@byu.edu} 
\author{Carsten M\"{u}ller}
\author{Karen Z. Hatsagortsyan}
\author{Christoph H. Keitel}
 \affiliation{Max-Planck-Institut f\"{u}r Kernphysik, Saupfercheckweg 1, D-69117
 Heidelberg, Germany}
\affiliation{$^{\dag}$Dept.\ of Physics and Astronomy,
Brigham Young University, Provo, UT 84602}

\date{\today}

\begin{abstract}
The radiation emitted by a single-electron wave packet in an intense
laser field is considered. A relation between the exact quantum
formulation and its classical counterpart is established via the electron's 
Wigner function. In particular we show that the wave packet, even when it 
spreads to the scale of the wavelength of the driving laser field, cannot be 
treated as an extended classical charge distribution but rather behaves as a
point-like emitter carrying information on its initial quantum state. 
We outline an experimental setup dedicated to put this conclusion 
to the test.
\end{abstract}

\pacs{41.60.-m, 42.50.Ct , 42.50.Xa }
\maketitle

The availability of super-intense lasers has stimulated interest in
relativistic electron dynamics in strong driving fields
\cite{Salamin2006Mourou2006}. Experimenters have observed the
effects of ponderomotive acceleration, the Lorentz drift, and
plasma wake-fields through direct detection of electrons ejected
from an intense laser focus. Photoemission from relativistically
driven plasmas has also been studied.

Much theory and computational effort has been devoted to the
dynamics of free-electron wave packets driven by intense fields
\cite{Roman2000Roman2001Mahmoudi2005,Mocken2005,Peatross2007} and the
associated scattered radiation 
\cite{Krekora2002,Chowdhury2005,Mocken2005}. Coherent
emission from many electrons can be viewed in the forward direction
with the emerging laser beam. Here, we consider
incoherent photoemission by free electrons out the side of a
focused laser, as a means of studying electron dynamics.

\begin{figure}[b]
    \centerline{\includegraphics[width=1.5in]{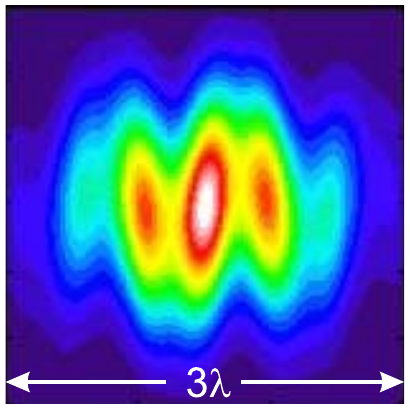}}
    \caption{An electron wave packet after natural spreading
    from an initially Gaussian-shaped size of 1{\AA}. The spreading takes 
    places during 190 cycles in a plane wave with intensity $2\times
    10^{18}~\mathrm{W/cm}^2$ and wavelength $\lambda=800$~nm.
    \label{fig1}}
\end{figure}

A free electron wave packet with an initial spatial size on the
scale of an atom undergoes natural quantum spreading, which eventually reaches
the scale of an optical wavelength, 
as illustrated in Fig.~\ref{fig1}
\cite{Peatross2007}. Moreover, an electron wave packet born through
field ionization is pulled from its parent atom at a finite rate,
typically emerging over multiple laser cycles. This, combined with
the Lorentz drift and sharp ponderomotive gradients found in a
tight relativistic laser focus, can cause a single-electron wave
packet to be strewn throughout a volume several laser wavelengths
across \cite{Chowdhury2005}; different portions of the same wave
packet can even be propelled out opposite sides of the laser focus.

The question naturally arises as to how a single-electron wave
packet radiates, especially when it undergoes such highly
non-dipole dynamics, where different parts of the electron wave
packet experience entirely different phases of a stimulating laser
field. So far, the problem has been treated \cite{Krekora2002,Chowdhury2005} within 
an intuitively appealing model where 
the quantum probability current is multiplied by the electron charge to produce an extended
current distribution used as a source in Maxwell's equations \cite{Jackson1998}. The intensity computed classically
from the extended current distribution is then associated with the
probability of measuring a photon. Due to interference, this approach 
can lead to dramatic suppression of radiation for many directions.
We note that semi-classical descriptions have, in general, proven very 
useful to understand processes like above-threshold ionization or 
high-harmonic generation, which arise from intense laser-matter 
interactions \cite{Corkum1993}.

In this letter, we provide a fully quantum mechanical treatment of
photoemission by a single-electron wave packet in a laser field
and relate it to a classical description via the electron's Wigner function. 
We show that no interference occurs between emission from different parts 
of an initially Gaussian wave packet, even if spatially large. 
In a plane-wave driving field, this result holds for wave packets of any shape 
and size.
The radiative response can be mimicked by the incoherent emission of a 
classical ensemble of point charges. 
In a focused laser beam, interferences can occur for certain initial 
electron states, but these interferences are of a different nature than 
those arising from a semi-classical current-distribution approach.
We outline an experimental arrangement able to probe the single-electron emission behavior by combining methods from strong-field physics and quantum optics.

\begin{figure}[htbp]
    \centerline{\includegraphics[width=2.5in]{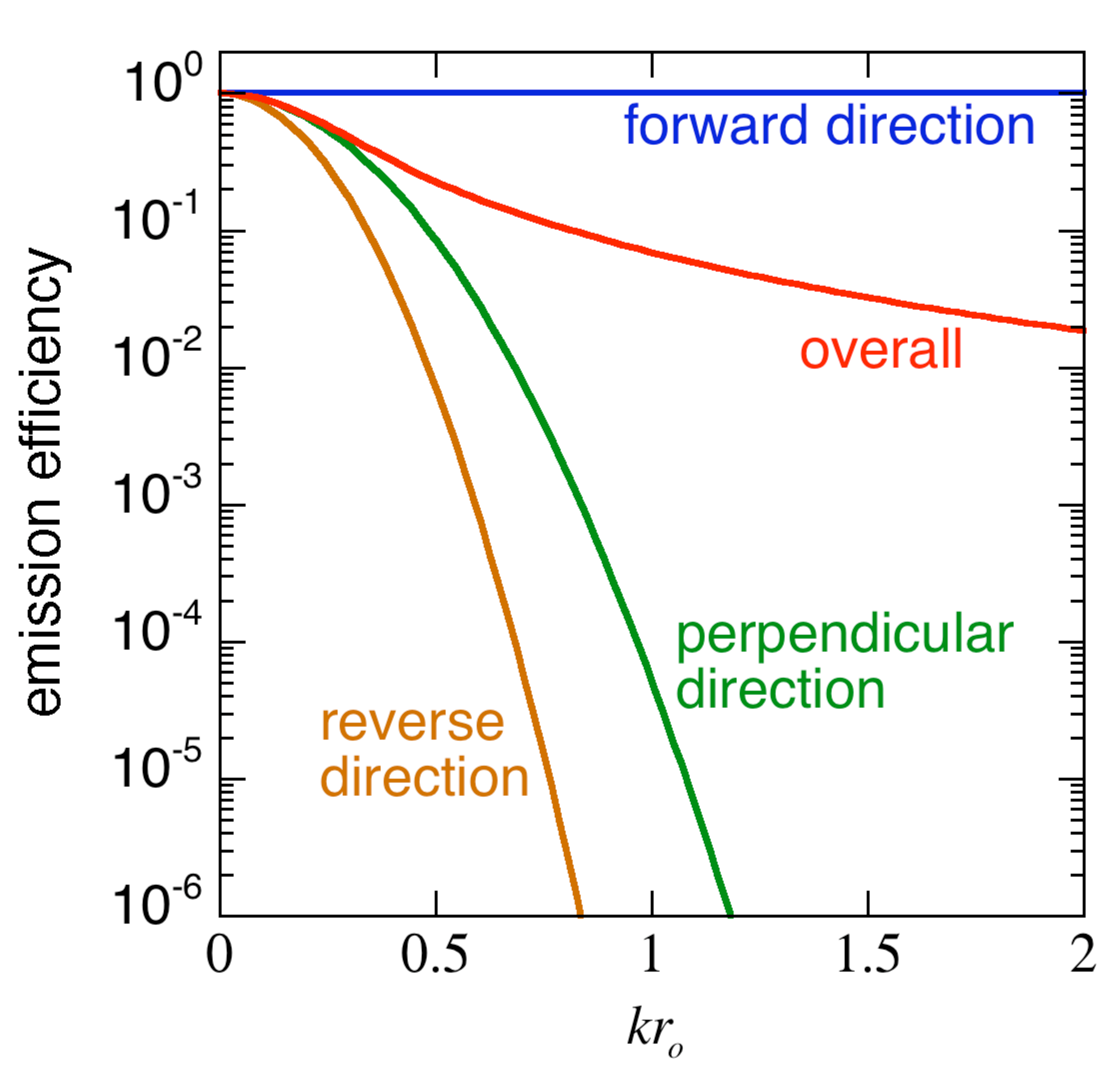}}
    \caption{Efficiency of light scattering into different
    directions as a function of size of a driven Gaussian charge
    distribution. \label{fig2} }
\end{figure}

We first examine radiation interferences that arise from treating a
single electron as an extended charge distribution in a semiclassical picture. 
Fig. 1 shows an example of an electron wave packet (probability density), computed
using the Klein-Gordon equation \cite{Peatross2007}. In the low-intensity limit, a wave-packet such as shown in Fig.~\ref{fig1} can be associated with a classical 
Gaussian current distribution: 
$\vec{J}\sim \hat {z}r_0^{-3} {\rm e}^{-{r^2} / {r_0^2 }}\,
{\rm e}^{i\kappa x}$, where $r_0$ characterizes the spatial
extent of the distribution with fixed overall charge. The
distribution is stimulated by a plane wave with wave four-vector 
$\kappa=(\omega_\kappa, \vecgreek{\kappa})$ traveling in the
$x$-direction and polarized in the $z$-direction. The intensity of the
Thomson scattering from the distribution is
\begin{equation} \label{eq1}
    I\left( {\theta ,\phi } \right)\sim \sin ^2\theta\, {\rm e}^{-|\vecgreek{\kappa}|^2 
    r_0^2 \left({1-\sin \theta \cos \phi } \right)},
\end{equation}
where $\theta =0$ defines the z-direction, and $\theta =\pi / 2$
with $\phi =0$ defines the x-direction. Fig.~\ref{fig2} shows the intensity
emitted into several directions, as well as the overall emission,
as a function of distribution size. The forward emission does not
vary from that of a single point oscillator with equal net charge.
In the perpendicular direction, the intensity drops by orders of
magnitude as the wave function grows to the scale of the wavelength
or bigger. This leads to a substantial loss in the overall
scattered-light energy \cite{near-field}.

Measurements of Compton/Thomson scattering provide an indication
that electrons do not radiate as extended
charge distributions. For example, $>$10~keV photons
scattered from electrons bound to helium corresponds to a scenario
where the size of the electron wave packet is larger than the
wavelengths involved. However, the cross section for the scattered
photons with energy well below the electron rest energy is known
\cite{Wollan1931,Jackson1998} to follow Eq.~(\ref{eq1}) with $r_0
=0$ (with $\phi $ averaged over all angles for unpolarized light).
It is interesting to note that A. H. Compton initially proposed a
``large electron'' model to explain the decrease in cross section
with angle for harder x~rays, which he later abandoned when the
effect of momentum recoil was understood \cite{Stuewer1975}.

Quantum electrodynamics provides the general framework to calculate
the radiation from a single-electron wave packet. The radiated
intensity is proportional to the expectation value of the
correlation function of the current density, rather than the
expectation value of the current density of the wave
packet \cite{Knight1980Sundaran1990Eberly1992}. From the
definition of the spectral intensity of the emitted radiation $d
\varepsilon_{\vec{k}} = \frac{c} {4\pi ^2} \left< \hat
{\vec{H}}_\omega ^{(-)} \hat {\vec{H}}_\omega ^{(+)} \right> R_0
^2d\Omega d\omega $, with radiation magnetic field $\hat
{\vec{H}}_\omega ^{(\pm )} =i\vec{k}\times \hat {\vec{A}}_\omega
^{(\pm )} $ and vector-potential $\hat {\vec{A}}_\omega ^{(\pm )}
=\frac{e{\rm e}^{ikR_0 }}{cR_0 }\int {d^4x\,\hat {\vec{J}}^{(\pm
)}(x){\rm e}^{ikx}} $, together with the current-density operators $\hat
{\vec{J}}^{(\pm )}$ in the Heisenberg representation, one derives
\begin{eqnarray} \label{eq2}
    d\varepsilon _{\vec{k}} &=& \frac{e^2}{4\pi ^2c}\int
    {d^4x} \int {d^4{x}'} {\rm e}^{ik(x-{x}')} \nonumber\\
& & \left< \left( {\vec{k}\times
    \hat {\vec{J}}^{(-)}(x)} \right)\left( {\vec{k}\times \hat {\vec
    {J}}^{(+)}({x}')} \right) \right> d\Omega d\omega .
\end{eqnarray}
Here $^{(\pm )}$ indicates the positive/negative frequency parts of
operators, $R_0$ the distance to the observation point from the
coordinate center, $\vec{k}$ the radiation wave-vector, $d\Omega$
the emission solid angle, $e$ the electron charge and $c$ the speed
of light. Equation~(\ref{eq2}) can be represented in a more
familiar form via the transition current density $\vec{J}_{{p}'i}
(x)$ in the Schr\"{o}dinger picture:
\begin{equation} \label{eq3}
    d\varepsilon _{\vec{k}} =\frac{e^2}{4\pi ^2c}\int {\frac{d^3{p}'}{(2\pi
    \hbar )^3}} \left| {\int {d^4x\,\left( {\vec{k}\times \vec{J}_{{p}'i} (x)}
    \right){\rm e}^{ikx}} } \right|^2d\Omega d\omega
\end{equation}
where $\vec{J}_{{p}'i} (x)=\left( {\bar {\psi }_{{\bf p}'}
(x)\vecgreek{\gamma}\psi _i^{({\rm L})} (x)} \right)$, 
$\psi _i^{({\rm L})} (x)$ is the initial
electron wave packet in the laser field, ${\psi }_{{\bf p}'}
(x)$ is a complete set of free electron states, and $\vecgreek{\gamma }$ are
the Dirac matrices. Equation~(\ref{eq3}) indicates that the total
probability of photon emission should be calculated as an
incoherent sum over the final momentum states of the electron, even
though in the experiment the final electron momentum could be
undetected.

Although Eq.~(\ref{eq3}) shows the general way to calculate the
emission intensity, it is difficult to apply in a real experimental
situation, as the quantum eigenstates of the electron in a focused
laser beam are usually unknown. It is therefore indeed desirable to
mimic the quantum electrodynamical result of Eq.~(\ref{eq3}) by
means of classical electrodynamical calculations (in the quasi-classical 
limit, when the photon energy is much smaller than the electron rest energy 
and recoil effects are negligible).
We demonstrate how to do this by way of the example of one-photon
Thomson scattering in a focused laser beam. Choosing the initial
wave packet in the form of $\psi _i (x)=\int {d^3p\,} \alpha
(\vec{p})\psi _{\vec{p}}(x)$, one can express the quantum-mechanical 
formula for spectral intensity of Eq.(\ref{eq3}) via the Wigner 
function $\rho_{\rm w} (\vec{r},\vec{p})=\int {d^3q\,
\alpha(\vec{p}+\vec{q}/2)\,} \alpha^\ast (\vec{p} - \vec{q}/2) {\rm e}^{i\vec{q}
\vec{r}}$ of the initial electron wave packet as follows:
\begin{equation} \label{eq4}
    d\varepsilon _{\vec{k}\lambda } =\frac{e^2\omega ^2}{4\pi ^2c^3}\int\!
    {d^3r\int\! {d^3p\,} \rho_{\rm w} (\vec{r},\vec{p})M_{\vec{k}\lambda } (\vec
    {r},\vec{p})} \,d\Omega d\omega ,
\end{equation}
where $M_{\vec{k}\lambda } (\vec{r},\vec{p})=(2\pi \hbar)^2 \int
{d^3} \kappa \,\int {d^3} {\kappa }'{\rm e}^{i(\vecgreek{\kappa
}-{\vecgreek{\kappa}}')\vecgreek{r}} \times A_0 (\vecgreek{\kappa })A_0^\ast ({\vecgreek{\kappa
}}')\Im _\lambda (\vec{p}_+ ,{\vec{p}}',\vecgreek{\kappa },\vec{k})\,\Im
_{_\lambda }^\ast (\vec{p}_- ,{\vec{p}}',{\vecgreek{\kappa
}'},\vec{k}) \delta (\varepsilon _{p_+} +\hbar \omega
_\kappa -\varepsilon _{{p}'} -\hbar \omega )\,\delta (\varepsilon
_{p_-} +\hbar \omega _{{\kappa }'} -\varepsilon _{{p}'} -\hbar
\omega )$ gives the radiation by an electron of momentum $\vec{p}$; 
$\Im _\lambda (\vec{p},{\vec{p}}',\vecgreek{\kappa
},\vec{k})$ is defined via $\int {d^4x\,} {\rm e}^{ikx}(\vec{e}_{_\lambda
}^\ast \vec{J}_{{p}'p} )=(2\pi \hbar )^4\int {d^3\kappa } A_0
(\vecgreek{\kappa })\Im _\lambda (\vec{p},\vec{{p}'},\vecgreek{\kappa
},\vec{k})\delta ^{(\ref{eq4})}(p+\hbar \kappa -{p}'-\hbar k)$,
$\varepsilon _p =c\sqrt {\vec{p}^2+m^2c^2} $ , $\vec{p}_\pm
=\vec{p}\pm \hbar ({\vecgreek{\kappa }'}-\vecgreek{\kappa })/2$ ,
${\vec{p}}'=\vec{p}+\hbar ({\vecgreek{\kappa }'}+\vecgreek{\kappa
})/2-\vec{k}$,
$\vec{e}_\lambda $ is the polarization of the emitted photon, and
$A_0 (\vecgreek{\kappa })$ the Fourier component of the focused laser
beam. When the Wigner function is non-negative
(e.g., for a Gaussian wave packet), 
it may be interpreted as the initial electron distribution in
phase space. The message of the structure of Eq.~(\ref{eq4}) is
that the total photoemission probability is an \textit{incoherent}
sum over the contributions of each local phase-space element of the
electron distribution. If, for example, the phase-space
distribution consists of two separate parts: $\rho_{\rm w} =
\rho_{\rm w}^{\rm (1)} + \rho_{\rm w}^{\rm (2)}$, then the intensities
emitted from each part incoherently add up to yield the total
radiation intensity. In the quasi-classical limit, the term
$M_{\vec{k}\lambda } \approx \left| {\mathcal{M}_{\vec{k}\lambda }
(\vec{r},\vec{p})} \right|^2$, with $\mathcal{M}_{\vec{k}\lambda }
(\vec{r},\vec{p})=2\pi \hbar \int {d^3\kappa \,{\rm e}^{i\vecgreek{\kappa
}\vecgreek{r}}A_0 (\vecgreek{\kappa })\Im _\lambda
(\vec{p},{\vec{p}}',\vecgreek{\kappa },\vec{k})} \,\delta (\varepsilon
_p +\hbar \omega _\kappa -\varepsilon _{{p}'} -\hbar \omega )$ can
be directly related to the classical electrodynamical calculation:
in the nonrelativistic limit $\Im _\lambda
(\vec{p},\vec{{p}'},\vecgreek{\kappa },\vec{k})=(\vec{e}_{_\lambda
}^\ast \vec{e}_\kappa )$ ($\vec{e}_\kappa$ is the polarization of
the $\vecgreek{\kappa}$-component of the driving field) and Eq.~(\ref{eq4})
describes the incoherent {\it average} of the radiation intensity over
the initial electron probability distribution in phase space. 
The radiation can thus be modelled by a classical ensemble of point emitters, taken individually.

A different situation arises when the Wigner function is negative in some phase-space region, indicating intrinsic quantum behavior.
As an example, we consider an electron in a superposition of two momentum states $|\vec{p}_1>$ and $|\vec{p}_2>$; in this case interference in the emitted radiation is possible. In fact, the final electron state $|\vec{p}^{\prime}>$ with emission of a photon of certain momentum $\vec{k}$ can be reached by two indistinguishable paths \cite{paths}: either from the state $|\vec{p}_1>$ or from $|\vec{p}_2>$ by absorption of different photons $\vecgreek{\kappa}_{1,2}=\vec{k}+\vec{p}'-\vec{p}_{1,2}$ from the external field, giving rise to interference. This effect is included in Eq.~(\ref{eq4})
which cannot be modelled by classical means here.
In the case of a Gaussian wave packet interferences are suppressed by the continuous spectrum of initial electron momenta which give rise to many interfering paths whose
contributions largely cancel out. In any case, interferences do not occur in a plane-wave driving field due to its uniform propagation direction $\vecgreek{\kappa}$. Then Eq.\,(\ref{eq4}) reduces to the incoherent superposition
$d\varepsilon _{\vec{k}\lambda } = \int\! {d^3p\,} |\alpha(\vec{p})|^2 
\varepsilon _{\vec{k}\lambda}(\vec{p})d\Omega d\omega$
of the spectral energies $\varepsilon _{\vec{k}\lambda}(\vec{p})= \frac{e^2\omega ^2}{4\pi ^2c^3} M_{\vec{k}\lambda}(0,\vec{p})$ radiated by an electron of momentum $\vec{p}$ \cite{spreading}.

The quantum interference effects above have to be clearly distinguished, though, from
those arising in the {\it coherent} part of the radiation mentioned in the introduction. The latter is calculated employing an ensemble average of the current operator $<\vec{J}(x)>$ as a source for the expectation value of 
the radiated field $<\vec{E}>$. In that picture, the spectral component of the coherent radiation intensity $|<\vec{E}_{\vec k}>|^2$ involves interference of the fields in the classical sense; i.e., all transitions with emission of a photon of certain momentum $\vec{k}$ interfere, independent of the final electron state. In the example of two superimposed momentum states, these are the transitions $|\vec{p}_i> \rightarrow |\vec{p}_j>$ with $i,j\in\{1,2\}$. The coherent radiation is analogous to the classical radiation of a modulated charge distribution here.

The coherent radiation is qualitatively different from the total radiation in Eq.~(\ref{eq4}). It accounts for the radiated field averaged over the quantum ensemble,
which excludes the incoherent field with a fluctuating phase
\cite{Marcuse1971}. In the case of a single electron, the coherent
radiation is only a small part of the total radiation, but becomes
dominant in the case of $N\gg 1$ radiating electrons which simultaneously 
interact with an applied field. As in the case
of radiation from an $N$-atom ensemble
\cite{Knight1980Sundaran1990Eberly1992}, the intensity of the
phase-matched coherent radiation is multiplied by a factor of
$N(N-1)$, while the incoherent radiation is multiplied by $N,
$ which becomes negligible. For example, any experiment on
high-harmonic generation from (many) atoms measures the coherent
emission (see, e.g., \cite{Lhuillier1992}). We note that the
terminology of a ``single-atom response'' that is commonly used in
this context, is therefore misleading.

Finally, we show that the radiation scattered from a single free electron 
in a laser is detectable by modern experimental techniques. The
feasibility of seeing scattered light depends crucially on whether
the electron wave packet radiates with the strength of a classical
point-like electron, as argued above. If one looks at light near
the frequency of the stimulating light (linear Thomson scattering),
the rate of emitted photons is proportional to the stimulating
intensity. Thus, beam fluence rather than peak intensity determines
the number of photons scattered by an electron in a laser.
Nevertheless, relativistic intensities may be helpful for
spectral-discrimination purposes: Because the Lorentz drift pushes the electron in the forward
direction of the laser, the scattered fundamental light is
typically red shifted from $\sim 800$~nm to longer wavelengths,
which is a signature that could be distinguished with bandpass
filters and detected with an avalanche photodiode. 

We envision an
electron born in the laser field (e.g. the second electron
pulled from He at $\sim 2 \times 10^{16}~\mathrm{W/cm}^2$).
Electrons given up by atoms during the leading edge of the pulse
tend to be pushed out of the focus by the ponderomotive gradient.
The density of donor atoms can be chosen such that on average one
electron experiences the highest intensities
($10^{-7}$~\mbox{torr}--$10^{-5}$~\mbox{torr}, depending on the
focal volume). Free electrons might also be prepared using a
suitable pre pulse. Only an electron experiencing the highest
intensity receives a substantial forward drift and the accompanying
red shift for emission in the direction perpendicular to laser
propagation. Keying in on this red-shift signature may be critical
for differentiating authentic scattered photon events from other
noise sources. Fast timing in the photon-detection electronics can
also suppress false signals, for example, scattered from the walls
of the experimental chamber.

\begin{figure}[b]
    \centerline{\includegraphics{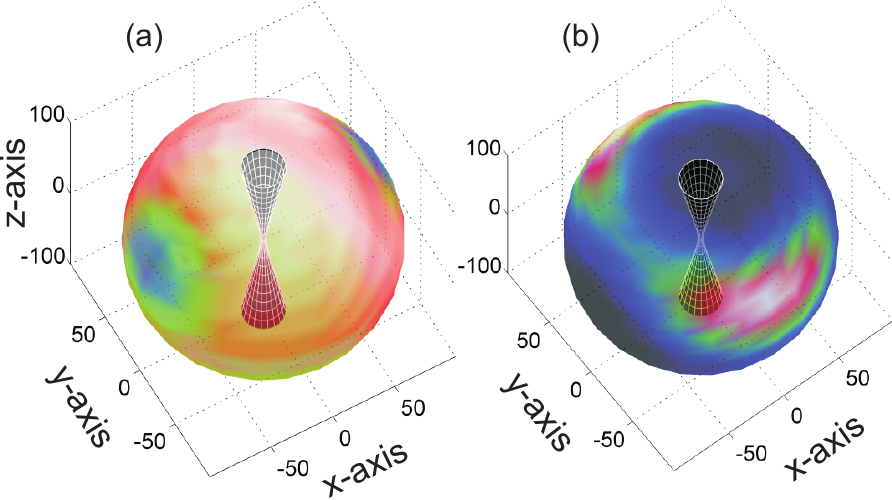}}
    \caption{ (a) Far-field intensity (at $100\mbox{ }\mu \mbox{m}$ 
    distance) of light scattered from a single
    electron trajectory born on axis during the early rising edge of an intense
    ($10^{19}$ W/cm$^2$) laser pulse. Red (blue) indicates regions of high (low)
    intensity. The laser beam (mesh) has waist $w_0=3\lambda $. The
    total scattered energy is 0.24~eV. (b) Far-field intensity 
    filtered to wavelengths between 850 nm and 950 nm with total 
    scattered energy 0.05 eV.\label{fig3}}
\end{figure}

We computed a representative single classical electron trajectory
in a tightly focused vector laser field with a peak intensity of
$10^{19}~\mathrm{W/cm}^2$, duration 35~fs, and wavelength 800~nm.
Fig.~\ref{fig3}(a) shows the total radiated energy in the far field emitted
from the electron as it is released on the rising edge of the
pulse. The electron trajectory eventually exits the side of the
focus due to ponderomotive gradients. Much of the scattered
radiation emerges out the side of the laser focus. The total
radiated energy (all angles and frequencies) from the single
electron trajectory is only $\sim0.24$~eV, indicating less than
one photon per shot.

 Fig.~\ref{fig3}(b) shows the spatial distribution
of light with wavelengths falling between 850~nm and 950~nm.
 This accounts for approximately 20{\%} of the total emitted power.
Assuming 10{\%} collection efficiency, this would amount to an
average 0.005~eV energy per shot, or one photon per 300 shots. This
can of course be increased if more electrons are used, where the
light radiated out the side of the focus adds incoherently for a
random distribution. The observation of blue-shifted light may also
afford an opportunity for discrimination against background. Intensities above $10^{19}~\mathrm{W/cm}^2$ are not ideal for this
experiment; a strong Lorentz drift redirects the photoemission into
the far forward direction.

In conclusion, we have studied the amount of light that an electron scatters 
out the side of a laser focus. 
We have shown that individual electrons radiate with the 
strength of point emitters. The electron's initial quantum state is imprinted
on the radiation spectrum via its Wigner function, which in general allows
for interference of different electron momentum components. The latter
is qualitatively distinct from the classical interference in the coherent
radiation of an extended charge distribution. Our results can be tested 
in an experiment that combines for the first time the sensitive techniques of 
quantum optics  (e.g., single-photon detectors) with the traditionally opposite
and incompatible discipline of high-intensity laser physics. 

The authors acknowledge Guido R. Mocken, J\"org Evers, Mikhail Fedorov, and Michael Ware for 
helpful input.


\begin{thebibliography}{99}

\bibitem{Salamin2006Mourou2006}
Y. I. Salamin {\it et al.}, Phys. Rep. {\bf 427}, 41 (2006);
G. A. Mourou, T. Tajima, and S. V. Bulanov, Rev. Mod. Phys. {\bf 78}, 309 (2006).

\bibitem{Roman2000Roman2001Mahmoudi2005}
J. San Rom\'{a}n, L. Roso, and H. R. Reiss, J. Phys. B {\bf 33}, 1869 (2000);
J. San Rom\'{a}n, L. Plaja, and L. Roso, Phys. Rev. A {\bf 64}, 063402 (2001);
M. Mahmoudi, Y. I. Salamin, and C. H. Keitel, {\it ibid.} {\bf 72}, 033402 (2005).

\bibitem{Mocken2005}
G. R. Mocken and C. H. Keitel, Comp. Phys. Comm. {\bf 166}, 171 (2005).

\bibitem{Peatross2007}
J. Peatross, C. M\"{u}ller, and C. H. Keitel, Opt. Expr. {\bf 15}, 6053 (2007).

\bibitem{Krekora2002}
P. Krekora {\it et al.}, Laser Phys. {\bf 12}, 455 (2002).

\bibitem{Chowdhury2005}
E. A. Chowdhury, I. Ghebregziabiher, and B. C. Walker,
J. Phys. B {\bf 38}, 517 (2005).

\bibitem{Jackson1998}
J. D. Jackson, {\it Classical Electrodynamics} (Wiley, New York, 1998), 3rd ed., Eq. (14.70); Eq. (14.124).

\bibitem{Corkum1993}
P. B. Corkum, Phys. Rev. Lett. {\bf 71}, 1994 (1993).

\bibitem{near-field}
Interferences in the radiation pattern from a
classical charge distribution as indicated by Fig.~\ref{fig2} imply a
corresponding amount of work exchanged between different portions
of the distribution via the near-field terms. The implication of
near-field work is problematic in the context of a single electron,
since one does not write a Hamiltonian for the interaction between
different parts of the same electron wave function. Note that
radiation reaction is negligible in the regime considered.

\bibitem{Wollan1931}
E. O. Wollan, Phys. Rev. {\bf 37}, 862 (1931).

\bibitem{Stuewer1975}
R. H. Stuewer, {\it The Compton Effect}, (Science History Publications,
New York, 1975).

\bibitem{Knight1980Sundaran1990Eberly1992}
P. L. Knight and P. W. Milonni, Phys. Rep. {\bf 66}, 21 (1980);
B. Sundaran and P. Milonni, Phys. Rev. A {\bf 41}, 6571 (1990);
J. H. Eberly and M. V. Fedorov, {\it ibid.} {\bf 45}, 4706 (1992).

\bibitem{paths} The two paths  leading to the same final states of the 
electron and the emitted photon are indistinguishable as the absorption 
of photons with different $\vecgreek{\kappa}_{1,2}$ from the external 
classical field are not detectable, even in principle, within the 
uncertainty of the driving coherent state.

\bibitem{spreading} As a consequence, in a plane-wave field Thomson scattering
from a Gaussian electron wave packet does not depend on the arrival time of the applied laser pulse, i.e., it does not depend on the size of the wave packet which freely spreads in the time preceeding the interaction.

\bibitem{Marcuse1971}
D. Marcuse, J. Appl. Phys. {\bf 42}, 2255 (1971).

\bibitem{Lhuillier1992}
A. L'Huillier {\it et al.}, Phys. Rev. A {\bf 46}, 2778 (1992);
W. Becker {\it et al.}, {\it ibid.} {\bf 56}, 645 (1997).

\end{thebibliography}

\end{document}